\documentclass{article}

\usepackage[utf8]{inputenc}
\usepackage[margin=1in]{geometry}
\usepackage[titletoc,title]{appendix}
\usepackage{url}
\usepackage{float}
\usepackage{ulem}   

\usepackage{amsmath,amsfonts,amssymb,mathtools}

\usepackage{graphicx,float}

\usepackage[ruled,vlined]{algorithm2e}
\usepackage{algorithmic}

\usepackage[autostyle]{csquotes}

 \usepackage[]{hyperref}

\title{An Introduction to Matrix factorization and Factorization Machines in Recommendation System, and Beyond}

\author{Yuefeng Zhang \\ yuefeng.zhang@pku.edu.cn}

\begin{document}

\date{}
\maketitle

\begin{abstract}

\noindent This paper aims at a better understanding of matrix factorization (MF), factorization machines (FM), and their combination with deep algorithms' application in recommendation systems. Specifically, this paper will focus on Singular Value Decomposition (SVD) and its derivations, e.g Funk-SVD, SVD++, etc. Step-by-step formula calculation and explainable pictures are displayed. What's more, we explain the DeepFM model in which FM is assisted by deep learning. Through numerical examples, we attempt to tie the theory to real-world problems.

\end{abstract}


\tableofcontents
\let\thefootnote\relax\footnote{We would like to thank \href{https://gaurav-khanna.com/about/}{Gaurav Khanna} for the valuable comments and help.}

\newpage

\section{Introduction and Overview}
General overview of MF and SVD algorithms' usage in the field of recommendation system.

\subsection{Problem Definition}

\noindent To start understanding matrix factorization for recommender systems, we define the following matrices: a rating matrix $R$ which is in general an $R \in \mathbb{R}^{m \times n}$ matrix.  Each row of $R$ represents a user and each column represents an item.  The values $R_{ui}$ represent the rating (or score) that user $u$ has given to item $i$.

\vspace{1\baselineskip}

\noindent The problem we are trying to solve is as follows:  For any collection of users and items, we have the situation where not all users have given ratings to all the items.  In fact, if examined the values of $R$, we would find that several values are incomplete.  However, we wish to recommend certain items to users.  How do we do that?  One way is to simply recommend any item the user has not already rated to them; this could mean sending a message to the user daily with a recommendation for an item.  However, we quickly see the issue with this; if we keep recommending items to users that they don't like, they will very quickly start ignoring our messages.  We would lose credibility with the user and they may eventually unsubscribe from our messages. 

\vspace{1\baselineskip}

\noindent So then we have to ask, ``What are the items a user {\it may} like?".  This is equivalent to asking, ``Of the items the user has not yet rated, what is our best guess for their ratings by a user?".  If we could come up with a system to guess this well, then we could send a daily email to the user recommending items which our system has predicted the user would rate highly.

\vspace{1\baselineskip}

\noindent Because the user will not rate all items, we expect missing values in this matrix. Our goal is to fill up those values. We use $\hat{R}$ to represent the predicted matrix according to our algorithm. 

\vspace{1\baselineskip}

\noindent There are two kinds of rating forms: implicit rating and explicit rating. Take movie recommendation as an example:

\begin{enumerate}

    \item Implicit: We only know the historical data whether the user watch the movie or not. The data can be represented as 0 and 1 format that 0 means un-watch and 1 means watch.
    
    \item Explicit: Users explicitly rate all the movie they watch in the past. The data can be represented in discrete range, e.g $1, 2, \cdots, 5$.
\end{enumerate}

\subsection{Matrix factorization (MF)}

\noindent One technique we can try is to use historical data to do recommendation. The most famous method for this problem is collaboration filter algorithm: neighborhood methods and Latent Factor Models (LFM). Matrix factorization is a successful implementation of the latter method. In fact, LFM and MF are talking about the same thing: how to auto-complete the score matrix by using dimension reduction methods.

\subsubsection{MF for different data sets}

\noindent There are two types of recommendation data set: explicit representing feedback data and implicit representing feedback data. The former one has score data while the later one only contains users' preference instead of numeric representation. So the implicit representing feedback data set only knows what users like but doesn't know what they hate. 

\vspace{1\baselineskip}

\noindent MF or LFM can perform great on the former data set because they are like machine learning's regression problem. For the later data set, we normally treat it as the top-$N$ recommendation problem and need reconstruct input data which can fit the algorithm. So we need to create negative data. There are two rules to construct negative samples for users:

\begin{enumerate}

    \item Make the number of positive and negative samples \textbf{balanced}.
    
    \item Prior to choose those \textbf{popular} items which aren't liked by the user. Because in this situation it's more likely for this user to unlike this particular popular item. The ratio of negative sample number to positive sample number usually set as $3$ in real problems. Experiments show that this ratio has effects in exploring recommendation system's long tail problem.

\end{enumerate}

\subsubsection{Comparison between MF and collaborative filtering algorithms}

\noindent Factorization machine and collaborative filtering algorithms are two major kinds of recommendation algorithms. There are two kinds of collaborative filtering algorithms: user-based and item-based. User-based collaborative filtering algorithm is to recommend the items which similar users have liked while item-based one is to recommend similar items according to users' part behaviour. Because internet products usually have much larger user number than item number, they prefer to use the item-based collaboration to the user-bases one in their recommendation system considering the computation complexity.

\noindent The biggest difference between MF and collaborative filtering (CF) algorithm is that the former one has the learning process and the latter one is based on statistics method. We are going to compare MF and collaborative filtering algorithms from the following aspects. Suppose we have $M$ users and $N$ items and the MF's latent factor dimension is $F$.

\begin{itemize}

    \item \textbf{Offline Computation Space Complexity}: For user-based CF, space complexity $O(M \cdot M)$; for item-based CF, it is  $O(N \cdot N)$. For MF , the space complexity is $O(F \cdot (M+N))$. Since $F$ is far less then $M$ or $N$, so MF method saves more memory,
    
    \item \textbf{Offline Computation Time Complexity}: Suppose we have $K$ records of users' rating on items. For user-based CF, time complexity is $O(N \cdot (N / K)^2)$; for item-based CF, it's $O(M \cdot (M / K)^2)$. Suppose we have $S$ iterations in MF, then its time complexity should be $O(S \cdot F \cdot K)$.
    
    \item \textbf{Online recommendation Time Complexity}: For CF algorithms, it can be done in real-time. In contrast, MF's recommendation process is time-consuming that its time complexity for generating a new recommendation result list is $O(F \cdot M \cdot N)$. So it's hard for MF method to do real-time recommendations. Nowadays, MF recommendation methods is usually divided into two steps: recall and rank. Recall is to select hundreds of potential items from the whole data set and rank is to finely rank those potential items into ordered recommendation list.

\end{itemize}


\section{Matrix factorization (MF)}

\subsection{Traditional SVD}

\noindent SVD\cite{billsus1998learning} was first used by Billsus and Pazzani in 1998 in recommendation service. They explored a few machine learning algorithms for collaborative filtering tasks and identified SVD as the best-performing one.

\vspace{1\baselineskip}

\noindent Eigendecomposition is used for square matrix and for non-square matrix we have SVD \cite{Eigen-SVD}. Assume we have $m$ users and $n$ items then score matrix will be $R \in \mathbb{R}^{m \times n}$. First, we complete missing values in score matrix as $R^{\prime}$, e.g. using global average value or user/item average value. Then, we use SVD to decompose $R^{\prime}$ as following\cite{billsus1998learning}:
\begin{equation}
R^{\prime} = U S V^{T}
\end{equation}
        
\noindent $U \in \mathbb{R}^{m \times m}$ and $V \in \mathbb{R}^{n \times n}$ are unitary matrices, $S \in \mathbb{R}^{k \times k}$ is diagonal matrix which $k \leq min(m, n)$ is the rank of $ R^{\prime}$ and values on its diagonal are the singular values of matrix $R$ in descending order.

\vspace{1\baselineskip}

\noindent We can try to understand SVD in the aspect of geometric transformation. Let's think matrix $R^{\prime}$ is transformation operation. The SVD factors one transformation matrix into three: (orthogonal) x (diagonal) x (orthogonal) which can be viewed from geometric aspect: (rotation) x (stretching) x (rotation). Figure shows this process. $USV ^T x$ starts with the rotation to $V^T x$. Then $S$ stretches that vector to $S V^T x$, and $U$ rotates to $Ax = U S V^T x$.

\begin{figure}[H]
    \centering
    \includegraphics[width=0.8\linewidth]{{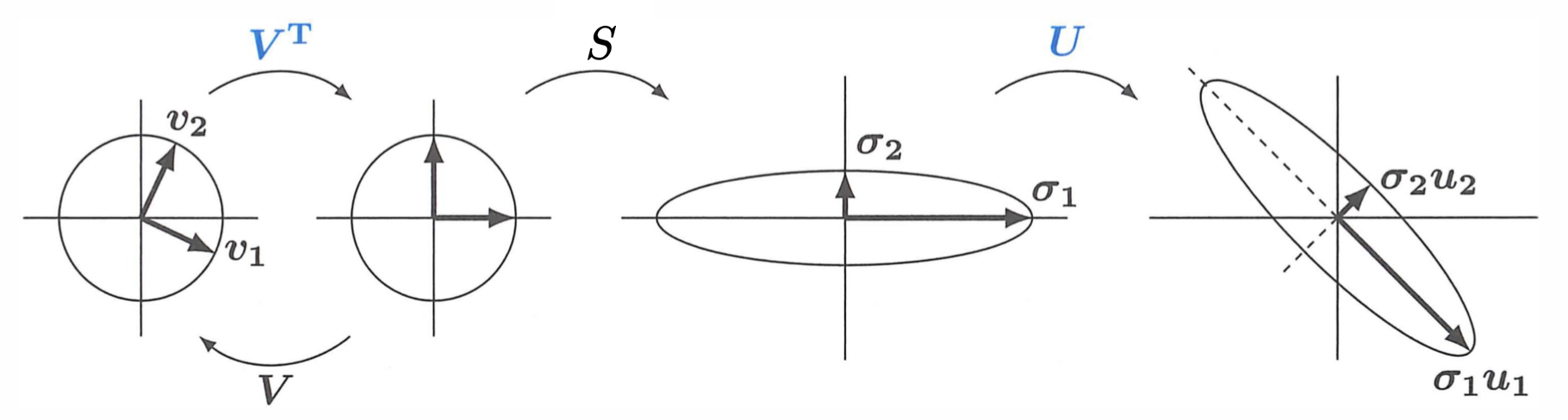}}
    \caption{SVD geometry explanation.}
    \label{fig:SVD-geometry}
\end{figure}

\vspace{1\baselineskip}

\noindent In order to do decomposition, we can choose the $f$ singular values to compose diagonal matrix $S_f$ and find corresponding rows and columns of those $f$ singular values in $U, V$ as $U_f, V_f$. Suppose the rank of $S$ is $r$. Only $r$ values in matrix $S$ are not empty. Values on $S$'s diagonal are in descending order which mean $s_{1,1} \leq s_{2,2} \leq \cdots \leq s_{r, r}$. We choose the $f$ by calculating the power of matrix $S_f$ to satisfy the equation \ref{equ:singular_power}. The $threshold$ is defined by ourselves and usually is set as $95\%$.
\begin{equation}
    \frac{\sum_{k} S_{k,k}^2}{\sum_i S_{i,i}^2} \times 100\% \geq threshold
    \label{equ:singular_power}
\end{equation}

\noindent Besides this heuristic method of choosing the rank $k$, there is a rule of thumb: choose $k$ such that the sum of the top k singular values is at least c times as big as the sum of the other singular values, where c is a domain-dependent constant (like 10, say). \cite{SVD-rule-choose-k}

\begin{equation}
\sum_{1}^f S_k \approx 10 \times  \sum_{f+1}^k S_k
\end{equation}

\vspace{1\baselineskip}

\noindent Then we can represent the score matrix after dimension decomposition as equation \ref{equ:SVD}.

\begin{equation}
    R_f^{\star} = U_f S_f V_f^{T}
    \label{equ:SVD}
\end{equation}

\begin{figure}[H]
    \centering
    \includegraphics[width=0.6\linewidth]{{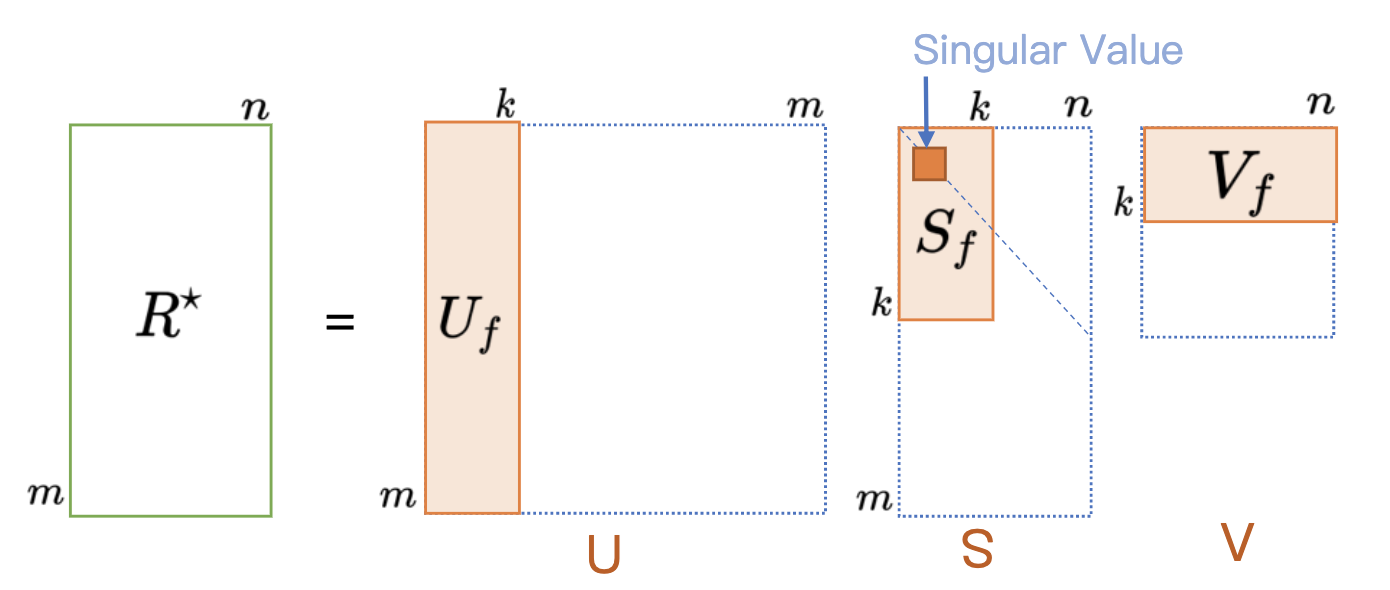}}
    \caption{How SVD does decomposition.}
    \label{fig:SVD}
\end{figure}

\noindent $R^{\star}_f(u, i)$ is the forecast rating score of user $u$ to item $i$ using decomposition representation. Equation \ref{equ:SVD} is the final representation of SVD usage in recommendation system. Figure \ref{fig:SVD} explains how SVD works. By using $U_f, S_f, V_f$ we can learn latent user's preference in an efficient way.

\noindent We can use SVD to decompose input matrix and utilize item-base collaboration method to recommend. The reason why we prefer item-base method to user-base method here is that for real applications user number is always larger than item number.

\vspace{1\baselineskip}

\noindent The $R_i$ column vector represents all of the ratings from users for item $i$.  Similarly, $R_j$ represents all of the users rating for item $j$.  We want to get a sense of whether items $i$ and $j$ are ``similar".  One way to measure this is to see how ``close" these vectors are to one another.  For example, if all the users gave both items $i$ and $j$ high ratings (i.e., ``4" or ``5:), then we would say that items $i$ and $j$ are similar.  From a vector point of view, these two vectors would likely be close to one another in $R^{1 \times n}$.  However, if all the users gave item $i$ a very high ratings (all ``5"s) but gave item $j$ very bad ratings (all ``1"s), then we might say that items $i$ and $j$ are not similar.  From a vector point of view, these two vectors would be far away from one another in $R^{1 \times n}$. 

\vspace{1\baselineskip}

\noindent To express the similarity in mathematical terms, we can utilize the dot product between two vectors. We can use Euclidean distance, Manhattan distance etc to measure the similarity. Here we choose the cosine similarity as the similarity metric. Concretely, we can define the {\bf cosine similarity} between two items as
\begin{equation}
    \cos (R_i, R_j)=\frac{R_i \cdot R_j}{\|R_i\|\|R_j\|}
\end{equation}

\vspace{1\baselineskip}

\noindent When calculating the similarity score between $R^{\star}_i$ and $R^{\star}_j$, for the item-based collaboration, we only consider those values which aren't empty in input rating matrix. So we define a mask matrix, $mask_{ij}$ which has the following values.

\begin{equation}
    \begin{aligned}
        mask_{ij} &= \begin{cases} 
                    1,  & \mbox{if }R_{i j} \mbox{ isn't empty} \\
                    0,  & \mbox{if }R_{i j} \mbox{ is empty}
                \end{cases} \\ 
    \end{aligned}
    \label{equ:mask}
\end{equation}

\noindent Essentially, this matrix places a value of 1 in positions where the original rating matrix, $R$, had a value; similarly, it places a value of zero where $R$ had a missing value.

\vspace{1\baselineskip}

\noindent For each empty value in rating matrix $R$, we calculate $\hat{R}_{u, i}$ based on user $u$'s rating of item $j$ (the rest of items) and the similarity score between item $i$ and item $j$. The process can be showed in equation \ref{equ:SVD_predict}. 
\begin{equation}
    \begin{aligned} 
        \mbox{similarity}_{total} &= \sum_{j \in n, j \neq i} \cos (R^{\star}_{i} \circ                                     mask_{i}, R^{\star}_{j} \circ mask_{j}) \\
        \hat{R}_{u, i} &= \sum_{j \in n, j \neq i} {\frac{\cos (R^{\star}_{i} \circ                      mask_{i}, R^{\star}_{j} \circ         mask_{,j})}{\mbox{similarity}_{total}} \cdot   R^{\star}_{j}}
    \end{aligned}
    \label{equ:SVD_predict}
\end{equation}

\noindent where $R_i$ is matrix $R$'s $i$th column vector, $\circ$ represents for the element-wise product (Hadamard product). Recall that the Hadamard Product between two matrices is defined as
\begin{equation}
\left[ A \circ B \right]_{ij} = \left[A \right]_{ij}  \left[B \right]_{ij} 
\end{equation}
for all $1 \leq i \leq m$ and $1 \leq j \leq n$. In order for this product to be defined, the matrices $A$ and $B$ have to be the same dimension but not necessarily square.  We should also emphasize that the end result of the Hadamard product is a matrix of the same size.  Therefore, the Hadamard product between two $\mathbb{R}^{1 \times n}$ vectors is also a $\mathbb{R}^{1 \times n}$ vector, and not a scalar as we would expect from a dot product.

\vspace{1\baselineskip}

\noindent Now we get the predicted matrix $\hat{R}$ by using SVD decomposition and item-base collaboration filter. The reason why we need the traditional SVD algorithm in recommendation is we want to do decomposition. Solving real world recommendation problems, we will face thousands and  millions of users and items to be recommended. It's a huge challenge for us to solve this problem in limited storage resources and computation resources. Matrix decomposition methods, e.g traditional SVD, help us to achieve the goal within those restrictions.

\vspace{1\baselineskip}

\noindent We may ask what do the $U$, $S$, and $V$ represent physically?  Do they have any physical significance?  For example each element in $R_{ui}$ represents the rating that user $u$ gave to item $i$.  Can we say the same for the elements in $U$ and $V$? The meaning of $U$, $S$, $V$ - they do not have a physical analogy like the matrix $R$. The values in the matrix can be negative, e.g the numeric example in section\ref{sec:Incomplete Matrix}, so that alone tells us that they are not representing any ``rating". But they can still be explained in an intuitive way. 

\vspace{1\baselineskip}

\noindent Every row of the matrix $U$ represents user's relationship with the particular category that every value means the category's importance (or relativity), the bigger the value is the more related they are. The matrix $U$'s column number is the category number we mention here and $U$'s row number is the number of users. Every column of the matrix $V$ represents the item's relationship with the particular category. The matrix $V$'s row number is the category number we mention here and $V$'s column number is the number of users. Matrix $S$'s values on the diagonal represent the relationship between users' category and items' category, the bigger the value is the more related they are. To better explain this part, especially what the category means, we will revisit this through the numerical example after equation \ref{equ:numeric-r-star}.

\subsubsection{Example - Complete Matrix}

Show example of how reducing dimsensionality (only selecting $f = 2$ can recreate an approximation of the original $R$.

\subsubsection{Example - Incomplete Matrix} 
\label{sec:Incomplete Matrix}

\noindent We'll first show an example of the power of SVD.

\vspace{1\baselineskip}

\begin{figure}[H]
    \centering
    \includegraphics[width=1\linewidth]{{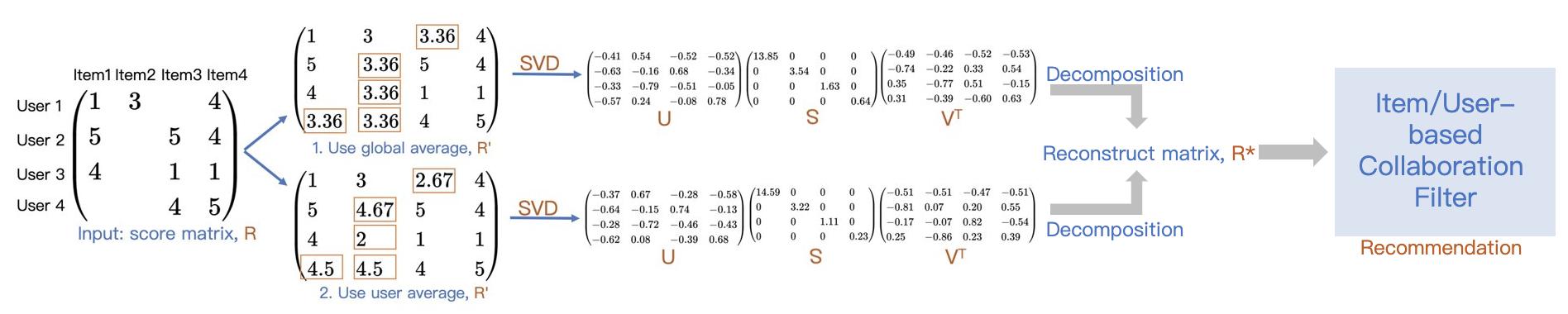}}
    \caption{SVD example.}
    \label{fig:SVD-example1}
\end{figure}

\noindent Here we take a small rank score matrix $R$ as an example to explain how this works visually. The rating system allows a user to give 1 - 5 discrete stars. As image \ref{fig:SVD-example1} shows, we have score matrix $R \in \mathbb{R}^{4 \times 4} $. Of the items the user has not yet rated, we use global average value or user/item average value to fill in. Then we use SVD on this already filled up matrix $R^{\prime}$.

\noindent Two different ways to complete the empty values:
\begin{enumerate}
    \item  \textbf{Use global average}: use the average value of all the existing values in rating matrix $R$. In our example, global average is calculated by 
    $$\frac{r_{1, 1} + r_{1, 2} + r_{1, 4} + r_{2, 1} + \cdots r_{4, 4}}{number_{non-empty}} = \frac{1+3+4+5+\cdots+5}{11} = \frac{37}{11} = 3.36$$
    
    \item  \textbf{Use user average}: use the average value of the existing ratings for the specific user by averaging across the rows for that user in matrix $R$. For user 1, the empty rating value can be calculated by 
    $$r^{\star}_{1,3} = \frac{r_{1, 1} r_{1, 2} + r_{1, 4}}{3} = \frac{1+3+4}{3} = 2.67$$
    
\end{enumerate}

\noindent Suppose we choose to use user average value to fill in empty values. We can set $f$ as 2. We can calculate the singular power percentage as equation \ref{equ:singular_power} which should be 
$$(14.59^2+3.22^2) / (14.59^2+3.22^2+1.11^2+0.23^2) \times 100\% = 99.42\%$$

\noindent Then the reconstructed matrix from rank 2 the $R^{\star}$ will be :
\begin{equation}
    \begin{aligned}
        R^{\star} &=U_{2} S_{2} V_{2}^{T} \\
        &=\left(\begin{array}{cccccccc}
        -0.37 & 0.67 \\
        -0.64 & -0.15 \\
        -0.28 & -0.72 \\
        -0.62 & 0.08
        \end{array}\right) \cdot\left(\begin{array}{cccc}
        14.59 & 0 \\
        0 & 3.22
        \end{array}\right) \cdot\left(\begin{array}{cccc}
        -0.51 & -0.51 & -0.47 & -0.51 \\
        -0.81 & 0.07 & 0.20 & 0.55
        \end{array}\right) \\
        &=\left(\begin{array}{cccc}
        0.98 & 2.87 & 2.96 & 3.88 \\
        5.14 & 4.7 & 4.32 & 4.46 \\
        3.94 & 1.88 & 1.45 & 0.76 \\
        4.39 & 4.60 & 4.33 & 4.71
        \end{array}\right)
    \end{aligned}
    \label{equ:numeric-r-star}
\end{equation}

\noindent Let's recall the SVD's physical explanation. First we want to be clear with a new concept category which means the latent classes we want to project users' information or items' information into. In the $SVD$ equation, the category for user is the column number of $U$ and the category for item is the column number of $V$. In $R^{\star}$'s reconstruction, we set the $f$ to 2 so the category we use to project the users' and items' latent information is 2. For example, the $1st$ row and $1th$ column value $-0.37$ of $U_2$ represents for user 1's relationship with user category 1. The the $1st$ row and $1th$ column value $0.98$ of $V_{2}^{T}$ represents for user 1's relationship with item category 1. And first value $14.59$ on $S_{2}$'s diagonal means the importance between user category 1 and item category 2.

\vspace{1\baselineskip}

\noindent Then we can combine the SVD decomposition with item-based collaboration to do predictions. Here we will explain how to predict one empty value detailedly. We can get the mask matrix according to equation \ref{equ:mask}.
\begin{equation}
    mask = \left(\begin{array}{llll}
               1&  1&   0&  1\\
               1&  0&   1&  1\\
               1&  0&   1&  1\\
               0&  0&   1&  1\\
            \end{array}\right)
\end{equation}

\noindent For the empty value $R_{3,2}$, we first calculate the total similarity score: 
\begin{equation}
    \begin{aligned}
    \text {similarity}_{\text {total}} &=\cos \left(R_{2}^{\star} \circ mask_{2}, R_{1}^{\star} \circ mask_{1}\right)+\cos \left(R_{2}^{\star} \circ mask_{2}, R_{3}^{\star} \circ mask_{3}\right)+\cos \left(R_{2}^{\star} \circ mask_{2}, R_{4}^{\star} \circ mask_{4}\right) \\
    &=\left(\begin{array}{c}
    2.87 \\
    0 \\
    0 \\
    0
    \end{array}\right) \cdot\left(\begin{array}{c}
    0.98 \\
    5.14 \\
    3.94 \\
    0
    \end{array}\right)+\left(\begin{array}{c}
    2.87 \\
    0 \\
    0 \\
    0
    \end{array}\right) \cdot\left(\begin{array}{c}
    0 \\
    4.32 \\
    1.45 \\
    4.33
    \end{array}\right)+\left(\begin{array}{c}
    2.87 \\
    0 \\
    0 \\
    0
    \end{array}\right) \cdot\left(\begin{array}{c}
    3.88 \\
    4.46 \\
    0.76 \\
    4.71
    \end{array}\right) \\
    &=2.8086+0+11.1356 \\
    &=13.94
    \end{aligned}
\end{equation}
\noindent We calculate the predicted value $\hat{R}_{3,2}$ as following:
\begin{equation}
    \begin{aligned}
         \hat{R}_{3,2} &= \frac{\cos(R^{\star}_2 \circ mask_2, R^{\star}_1 \circ        mask_1) R^{\star}_{3, 1} + 
                         \cos(R^{\star}_2 \circ mask_2, R^{\star}_3 \circ mask_3) R^{\star}_{3, 3} + 
                         \cos(R^{\star}_2 \circ mask_2, R^{\star}_4 \circ mask_4) R^{\star}_{3, 4}
                         }{\mbox{similarity}_{total}} \\
                      &= \frac{3.94 \times 2.81 + 0 + 0.76 \times 11.14}{13.94} \\
                      &= 19.5289 / 13.94 \\
                      &= 1.4
    \end{aligned}
    \label{equ:}
\end{equation}

\noindent So we gain the predicted value $\hat{R}_{3,2}$. If we want to use discrete ratings $1, 2, \cdots, 5$ we can take the value 1.4 in round to 1. The rest missing values can be predicted in the same manner.

\vspace{1\baselineskip}

\subsubsection{Example - Numerical Decomposition of Matrices}

\begin{equation}
\chi^2 = (1 - l_{11} u_{11} + l_{12} u_{21})^2 + (3 - l_{11} u_{12} + l_{12} u_{22}) + ...
\end{equation}
                      
\subsection{Funk-SVD}

\noindent Traditional SVD needs large storage for completed matrix and its computation complexity is high. In order to solve those two shortcomings, Simon Funk proposed Funk-SVD\cite{Funk-SVD} in 2006 during the ongoing of Netflix Prize competition. Funk-SVD isn't based on heuristic similarity but applies explicit learning process. Funk-SVD can also be called by latent factor model (LFM).

\vspace{1\baselineskip}

\noindent As we saw in our SVD example, we want to take advantage of an important fact: that the matrices that result in the decomposition can have significantly lower dimension than the original matrix.  Here, $f$ is chosen arbitrarily and is a smaller number compared with $m$ and $n$. However, unlike SVD, we do not make any assumptions about the character of the $P$ and $Q$ matrices in $ \hat{R} = P^{T} Q$.  That is, in SVD we knew that $U$ and $V$ were unitary matrices and $S$ was a diagonal matrix.  In general, for $A = LU$ decomposition, we know that $L$ and $U$ are upper-triangular and lower-triangular matrices, respectively.  In the Funk SVD process, we do not place any expectations on $P$ and $Q$; they are simply two component matrices whose product will approximate the original rating matrix.

\vspace{1\baselineskip}

\noindent Let's define score matrix $R \in \mathbb{R}^{m \times n}$, where $m$ is the number of users and $n$ is the number of items. Each value in this matrix represents user's rating on item, e.g $\hat{R}(u, i)$ is user $u$'s rating on item $i$. Because we can tens of thousands (if not more) of different items in a real-world scenario, so the dimension of $\hat{R}$ can have problems in computation efficiency. So based on the idea of SVD, Funk proposed this algorithm to compute users' preference more efficiently.

\vspace{1\baselineskip}

\noindent As we saw with SVD, we can decompose $R$ into three matrices.  In the Funk-SVD process, we can decompose score matrix $R$ into the multiplication of two low-dimensional matrices, $P \in \mathbb{R}^{f \times m}$ and $Q \in \mathbb{R}^{f \times n}$\cite{Funk-SVD}. 

\vspace{1\baselineskip}

\noindent As we saw in our SVD example, we want to take advantage of an important fact: that the matrices that result in the decomposition can have significantly lower dimension than the original matrix.  Here, $f$ is chosen arbitrarily and is a smaller number compared with $m$ and $n$. However, unlike SVD, we do not make any assumptions about the character of the $P$ and $Q$ matrices in $ \hat{R} = P^{T} Q$.  That is, in SVD we knew that $U$ and $V$ were unitary matrices and $S$ was a diagonal matrix.  In general, for $A = LU$ decomposition, we know that $L$ and $U$ are upper-triangular and lower-triangular matrices, respectively.  In the Funk SVD process, we do not expect the same for $P$ and $Q$.

\vspace{1\baselineskip}

\begin{figure}[H]
    \centering
    \includegraphics[width=0.7\linewidth]{{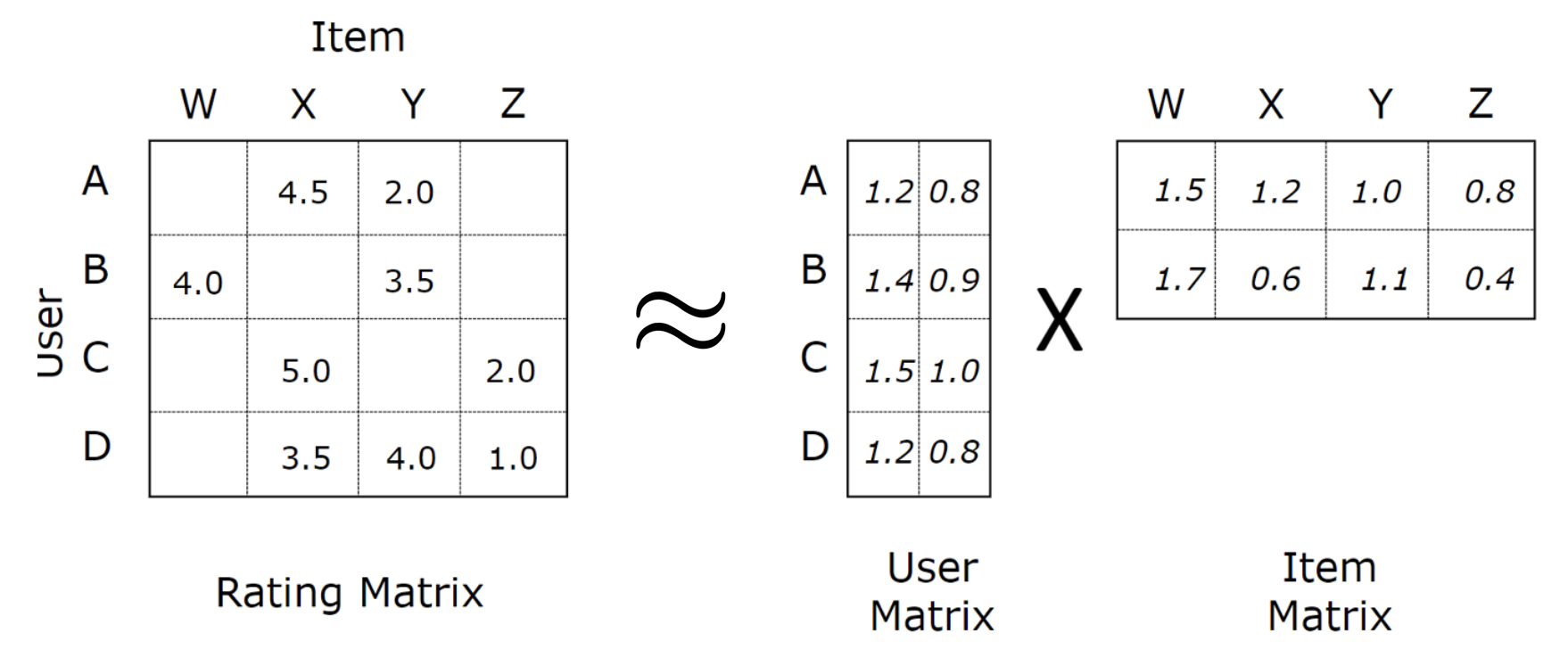}}
    \caption{How Funk-SVD works.}
    \label{fig:PQMulti}
\end{figure}

\noindent The mathematics of Funk-SVD can be written as follows:

\begin{equation}
    \begin{aligned}
        \hat{R}         &= P^{T} Q \\
        \hat{R}(u, i)  &=\hat{r}_{u i} \\
        \hat{r}_{u i}   &=\sum_{f} p^{T}_{u f} q_{i f}
    \end{aligned}
\end{equation}
where $\hat{r}_{u i}$ is the predicted value which can be calculated by $P$ and $Q$ matrices. So the input rating matrix $\hat{R}$ can have many empty value in it and we use the multiplication of $P$ and $Q$ to represent it. Then we can predict those empty values which are our interested parts though the multiplication result of $P$ and $Q$. 
By using a simple example as figure \ref{fig:PQMulti}, we can figure out how this decomposition and prediction process works.

\noindent In this example, our input rating matrix $R \in \mathbb{R}^{4 \times 4}$ has four users and fours items and we use the multiplication result of user matrix $P^{T} \in \mathbb{R}^{4 \times 2}$ and item matrix $Q \in \mathbb{R}^{2 \times 4}$ to represent it. For the already existing value $R_{A Y} = 2.0$ is nearly represented by $P^{T}_{A} \times Q_{Y} = 1.2*1.0 + 0.8*1.1 = 2.08$. After obtaining those two decomposition matrices $P$, $Q$ we can predict the missing values in the rating matrix from $P$, $Q$'s product. In our example, the missing value $R_{A Z}$ can be gained by $P^{T}_{A}*Q_{Z} = 1.2*0.8 + 0.8*0.4 = 1.28$.

After mastering the whole process of Funk-SVD, we can focus on how to get $P$ and $Q$. Not like the traditional SVD algorithm, we can apply implicit learning process to get those two matrices, i.e we can set the aim to minimum cost on training data to learn $P, Q$ matrices. The loss function $C(p, q)$ contains two parts: squared loss and regularization.
\begin{equation}
    C(p, q)=\sum_{(u . i) \in \text { Train }}\left(r_{u i}-\sum_{f=1}^{F} p_{u f} q_{i f}\right)^{2}+\lambda\left(\left\|p_{u}\right\|^{2}+\left\|q_{i}\right\|^{2}\right)
\end{equation}
Here $\lambda$ is the regularization parameter. Stochastic gradient descent (SGD) and other optimization algorithms can be applied to learn the minimum loss and its corresponding $P, Q$. We have two target parameters here $p_{u f}$ and $q_{i f}$, so we first calculate $C(p, q)$'s partial derivation over them:
\begin{equation}
    \begin{aligned}
        \frac{\partial C}{\partial p_{u k}}= \sum_{i}{2(r_{u i}- \sum_f{p_{u f}q_{i f}} )}(-q_{i k}) + 2 \lambda p_{u k} \\
        \frac{\partial C}{\partial q_{i k}}= \sum_{u}{2(r_{u i}- \sum_f{p_{u f}q_{i f}} )}(-p_{u k})+2 \lambda q_{i k}
    \end{aligned}
\end{equation}
Here we use $k$ to represent the $f$ variable. For each $u$ or $i$, define $err_{u i} = r_{u i}- \sum_f{p_{u f}q_{i f}}$. We will talk about the details of SGD and other optimization functions in section \ref{sec:optimization}. And according to SGD, we update parameters as equation \ref{equ:FUNK-SGD} for each $i$ or $u$ in the loop.
\begin{equation}
    \begin{aligned}
        p_{u k}=p_{u k}+\alpha\left(q_{i k} \cdot err_{u i} - \lambda p_{u k}\right) \\
        q_{i k}=q_{i k}+\alpha\left(p_{u k} \cdot err_{u i} - \lambda q_{i k}\right)
    \end{aligned}
    \label{equ:FUNK-SGD}
\end{equation}
where $k \in [0, f]$, $\alpha$ is the learning rate. The pseudo-code of Funk-SVD is as algorithm \ref{alg:Funk-SVD}.

\begin{algorithm}[H]
    \begin{algorithmic}
        \STATE{\textbf{Initialize $P, Q$}}:
            \STATE{$p_{i k}, q_{i, k} \gets random() / sqrt(f)$ }
        \FOR{$step = 1:n$}
            \FOR{$u, i, r_{u i}$ in train.items()}
                \STATE{$err_{u i} \gets r_{u i} -  <p[u, :f+1], q[i, :f+1]>$}
                \STATE{\textbf{Update {p, q}:}}
                \FOR{$k = 1:f$}
                    \STATE{$p_{u k} += \alpha \cdot(q_{i k} \cdot err_{u i}-\lambda p_{u k})$}
                    \STATE{$q_{i k} += \alpha \cdot(p_{u k} \cdot err_{u i}-\lambda q_{i k})$}
                \ENDFOR
            \ENDFOR
        \ENDFOR

        \RETURN{$P, Q$}

    \end{algorithmic}
    \caption{Funk-SVD Algorithm}
    \label{alg:Funk-SVD}
\end{algorithm}

\vspace{1\baselineskip}

\noindent There are many different implements of this algorithm and we follow what Funk did in his original paper. The logic behind why this implement update each $k$ dimension one by one, quote from Funk: 
\blockquote{``Anyway, this will train one feature (aspect), and in particular will find the most prominent feature remaining (the one that will most reduce the error that's left over after previously trained features have done their best). When it's as good as it's going to get, shift it onto the pile of done features, and start a new one. For efficiency's sake, cache the residuals (all 100 million of them) so when you're training feature 72 you don't have to wait for predictRating() to re-compute the contributions of the previous 71 features. You will need 2 Gig of ram, a C compiler, and good programming habits to do this."}

\vspace{1\baselineskip}

\subsection{SVD++}
SVD++\cite{koren2010factor} is raised by Yehuda Koren etc. in Netflix Prize competition. It includes users' historical scored items into LFM models which takes neighbour effects into consideration. SVD++, like Funk-SVD, learns neighbour relationship by minimize a global loss.

SVD++\cite{koren2010factor} is the explicit learning improvement of Item-to-Item Collaboration Filtering (ItemCF) algorithm \cite{linden2003amazon}. Let's first review how ItemCF works.
ItemCF first calculates the similarity between items which is represented as $w_{i,j}$ and then uses item similarity score and users' historical behaviour to fill up score matrix. Assume $p_{uj}$ is user $u$'s score on item $j$. ItemCF calculated $p_{uj}$ as equation \ref{equ:ItemCF}.
\begin{equation}
    p_{u j}=\sum_{i \in N(u) \cap S(j, K)} w_{j i} r_{u i}
    \label{equ:ItemCF}
\end{equation}
where $N(u)$ is user $u$'s rated item set, $S(j, K)$ is set of top-K most related items to item $j$, $r_{u i}$ is user $u$'s score on item $i$.
In ItemCF $w_{i,j}$ is defined by $\displaystyle w_{i j}=\frac{|N(i) \cap N(j)|}{|N(i)|}$ where $|N(i)|, |N(j)|$ are the number of users who are interested in item $i, j$, but in SVD++ $w_{i j}$ is a learnable parameter. 
Based on ItemCF's idea and assume $w_{i j}$ is a learnable parameter, we can write the SVD++ equation as \ref{equ:SVD++_1}. SVD++ can be seen as an extension of Funk-SVD to implicit rating value.
\begin{equation}
    \hat{r}_{u i}=\frac{1}{\sqrt{|N(u)|}} \sum_{j \in N(u)} w_{i j}
    \label{equ:SVD++_1}
\end{equation}
Where $\hat{r}_{u i}$ is the predicted score of user $u$ on item $i$. Like Funk-SVD algorithm, loss object of SVD++ can be defined as following equation:
\begin{equation}
    C(w)=\sum_{(u, i) \in \text { Train }}\left(r_{u i}-\sum_{j \in N(u)} w_{i j} r_{u j}\right)^{2}+\lambda w_{i j}^{2}
\end{equation}
where $\lambda$ is the regularization parameter. Because $w_{i j}$'s number of parameters can be very large, so Koren proposed to do dimensionality reduction on $w$. We can decompose $w$ into two low-dimension vectors' product. Then equation \ref{equ:SVD++_1} can be rewritten as following equation\ref{equ:SVD++_2}.
\begin{equation}
    \begin{aligned}
        \hat{r}_{u i}=\frac{1}{\sqrt{|N(u)|}} \sum_{j \in N(u)} x_{i}^{T} y_{j}\\
                     =\frac{1}{\sqrt{|N(u)|}} x_{i}^{T} \sum_{j \in N(u)} y_{j}
    \label{equ:SVD++_2}
    \end{aligned}
\end{equation}
where $x_i, y_j$ are two F-dimension vectors, so $w$'s number of parameters is eliminated from $O(i\cdot j) $to $O(2\cdot n\cdot F)$. Moreover, we can add bias terms into equation \ref{equ:SVD++_2} and combine it with Funk-SVD \ref{alg:Funk-SVD}.
\begin{equation}
    \hat{r}_{u i}=\mu+b_{u}+b_{i}+p_{u}^{T} \cdot q_{i}+\frac{1}{\sqrt{|N(u)|}} x_{i}^{T} \sum_{j \in N(u)} y_{j}
\end{equation}
where $\mu$ is global average score of the rating matrix, $b_{u}$ is user bias, $b_{i}$ is item bias. Those two bias vectors $b_{u}$ and $b_{i}$ are learnable parameters.
Also, in order to not have too many parameters to cause over-fitting, we can arbitrarily set $x = q$ to reduce the number of parameters. So the final SVD++ can be written as the following equation\ref{equ:SVD++}.
\begin{equation}
    \hat{r}_{u i}=\mu+b_{u}+b_{i}+q_{i}^{T} \cdot (p_{u}+\frac{1}{\sqrt{|N(u)|}} \sum_{j \in N(u)} y_{j})
    \label{equ:SVD++}
\end{equation}
The learning process of parameters $b_u, b_i, p, q$ is similar to Funk-SVD. We can apply SGD according to their partial derivations. 

\section{Factorization Machine (FM)}
Factorization Machine (FM)\cite{FM} is designed to combine Support Vector Machine (SVM) and MF. It can handle large sparse data and subsumes MF and SVD++. FM does not only use first-order feature linear components but also second-order (cross-product) of the features to capture more potential relationship inside the features.

Suppose we have $x$ as input matrix (rating and other features) and $\hat{y}$ as output. 2-way FM can be written as equation\ref{equ:FM} which only concerns first-order and second-order feature interactions. An example of the input and output data format as figure.
\begin{figure}[H]
    \centering
    \includegraphics[width=0.8\linewidth]{{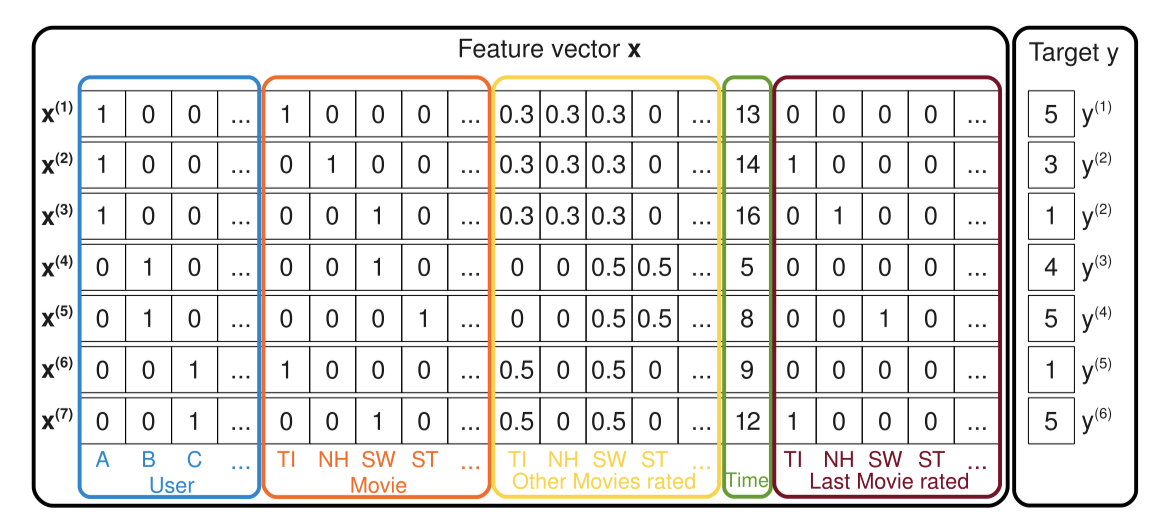}}
    \caption{Example of FM's input and output data format.}
    \label{fig:FM}
\end{figure}

\begin{equation}
    \begin{aligned}
        \hat{y}(\mathbf{x}) = w_{0}+\sum_{i=1}^{n} w_{i} x_{i}+\sum_{i=1}^{n} \sum_{j=i+1}^{n}\left\langle\mathbf{v}_{i}, \mathbf{v}_{j}\right\rangle x_{i} x_{j}\\
        \left\langle\mathbf{v}_{i}, \mathbf{v}_{j}\right\rangle = \sum_{f=1}^{k} v_{i, f} \cdot v_{j, f}
    \end{aligned}
    \label{equ:FM}
\end{equation}
where we have three parameters: $w_{0} \in \mathbb{R}$, $\mathbf{w} \in \mathbb{R}^{n}$, $\mathbf{V} \in \mathbb{R}^{n \times k}$ that $w_{0}$ is the global bias, $w_i$ models the strength of the i-th variable, $\hat{w}_{i, j}:=\left\langle\mathbf{v}_{i}, \mathbf{v}_{j}\right\rangle$ models the interaction between the ith and j-th variable which idea is similar with FM that can do dimensionality reduction on $w$. 
We can prove the computation complexity of equation \ref{equ:FM} is linear.
\begin{equation}
    \begin{aligned}
        & \sum_{i=1}^{n} \sum_{j=i+1}^{n}\left\langle\mathbf{v}_{i}, \mathbf{v}_{j}\right\rangle x_{i} x_{j} \\
        =& \frac{1}{2} \sum_{i=1}^{n} \sum_{j=1}^{n}\left\langle\mathbf{v}_{i}, \mathbf{v}_{j}\right\rangle x_{i} x_{j}-\frac{1}{2} \sum_{i=1}^{n}\left\langle\mathbf{v}_{i}, \mathbf{v}_{i}\right\rangle x_{i} x_{i} \\
        =& \frac{1}{2}\left(\sum_{i=1}^{n} \sum_{j=1}^{n} \sum_{f=1}^{k} v_{i, f} v_{j, f} x_{i} x_{j}-\sum_{i=1}^{n} \sum_{f=1}^{k} v_{i, f} v_{i, f} x_{i} x_{i}\right) \\
        =& \frac{1}{2} \sum_{f=1}^{k}\left(\left(\sum_{i=1}^{n} v_{i, f} x_{i}\right)\left(\sum_{j=1}^{n} v_{j, f} x_{j}\right)-\sum_{i=1}^{n} v_{i, f}^{2} x_{i}^{2}\right) \\
        =& \frac{1}{2} \sum_{f=1}^{k}\left(\left(\sum_{i=1}^{n} v_{i, f} x_{i}\right)^{2}-\sum_{i=1}^{n} v_{i, f}^{2} x_{i}^{2}\right)
    \end{aligned}
\end{equation}
So equation \ref{equ:FM}'s computation complexity is  $ O(k \cdot n)$ where $k$ is latent factor's dimension and $n$ is input $x$'s dimension. We can compute those model parameters with gradient decent methods for a variety of losses. The partial gradient of parameters in equation \ref{equ:FM} can be computed as following:
\begin{equation}
    \frac{\partial}{\partial \theta} \hat{y}(\mathbf{x})=\left\{\begin{array}{ll}
    1, & \text { if } \theta \text { is } w_{0} \\
    x_{i}, & \text { if } \theta \text { is } w_{i} \\
    x_{i} \sum_{j=1}^{n} v_{j, f} x_{j}-v_{i, f} x_{i}^{2}, & \text { if } \theta \text { is } v_{i, f}
    \end{array}\right.
\end{equation}
\enquote{In practice, usually a two-way FM model is used, i.e., only the second-order feature interactions are considered to favor computational efficiency.}\cite{MS-FM}
The equation for the multi-way feature interactions can be found in the original paper\cite{FM}.

\subsubsection{Field-Aware Factorization Machine (FFM)}
\noindent Field-aware factorization machine (FFM) \cite{juan2017field} is an extension to FM. For traditional recommendation problems, e.g. movie recommendation, there are only three available fields User, Item, and Tag. But when it comes to click-through-rate prediction problem, we can have multiple fields, e.g. Publisher (P), Advertiser (A), Gender (G), Click Result. We recognize those feature categories as \textit{fields}. FFM is the upgrade of factorization machine (FM) by adding the field concept: it considers similar features as a same field. The advantage of FFM over FM is that, it uses different factorized latent factors for different fields of features. \enquote{Putting features into fields resolves the issue that the latent factors shared by features that intuitively represent different categories of information may not well generalize the correlation.}\cite{MS-FM}

\vspace{1\baselineskip}

\noindent Suppose $\mathbf{X} \in \mathbb{R}^{n \times k}$, $k$ is latent factor's dimension and $f$ is the number of fields. FFM equation based on 2-way FM equation can be written as following:
\begin{equation}
    \theta_{\mathrm{FFM}}(\mathbf{w} \mathbf{x})=\sum_{j_1 = 1}^{n} \sum_{j_2=j_ 1+1}^{n} <\mathbf{v}_{j_1, f_2} , \mathbf{v}_{j_2,f_1}> x_{j_1} x_{j_2}
    \label{equ:FFM}
\end{equation}

\noindent where $f_1$ and $f_2$ are respectively the ﬁelds of $j_1$ and $j_2$ . FFM's number of variables is $O(f\cdot n \cdot k)$, equation \ref{equ:FFM}'s computation complexity is $O(n^2 k)$.

\subsection{With Deep Learning}
\subsubsection{DeepFM}
\noindent Factorization-Machine based neural network(DeepFM) is a evolutional model of Wide \& Deep model from Google. DeepFM has a share input of its 'wide' and 'deep' part which behind idea is quit simliar with the traditional CTR model, Factorization Machine(FM). The model aims at predicting CTR by exphasizing both low- and high- order feature interactions.

\vspace{1\baselineskip}

\noindent DeepFM has two components: FM component and deep component. The input of these two components is the same embedding of input features. The output of the model is the sigmoid result of the combination of this two components. 

\vspace{1\baselineskip}

\begin{figure}[H]
    \centering
    \includegraphics[width=0.8\linewidth]{{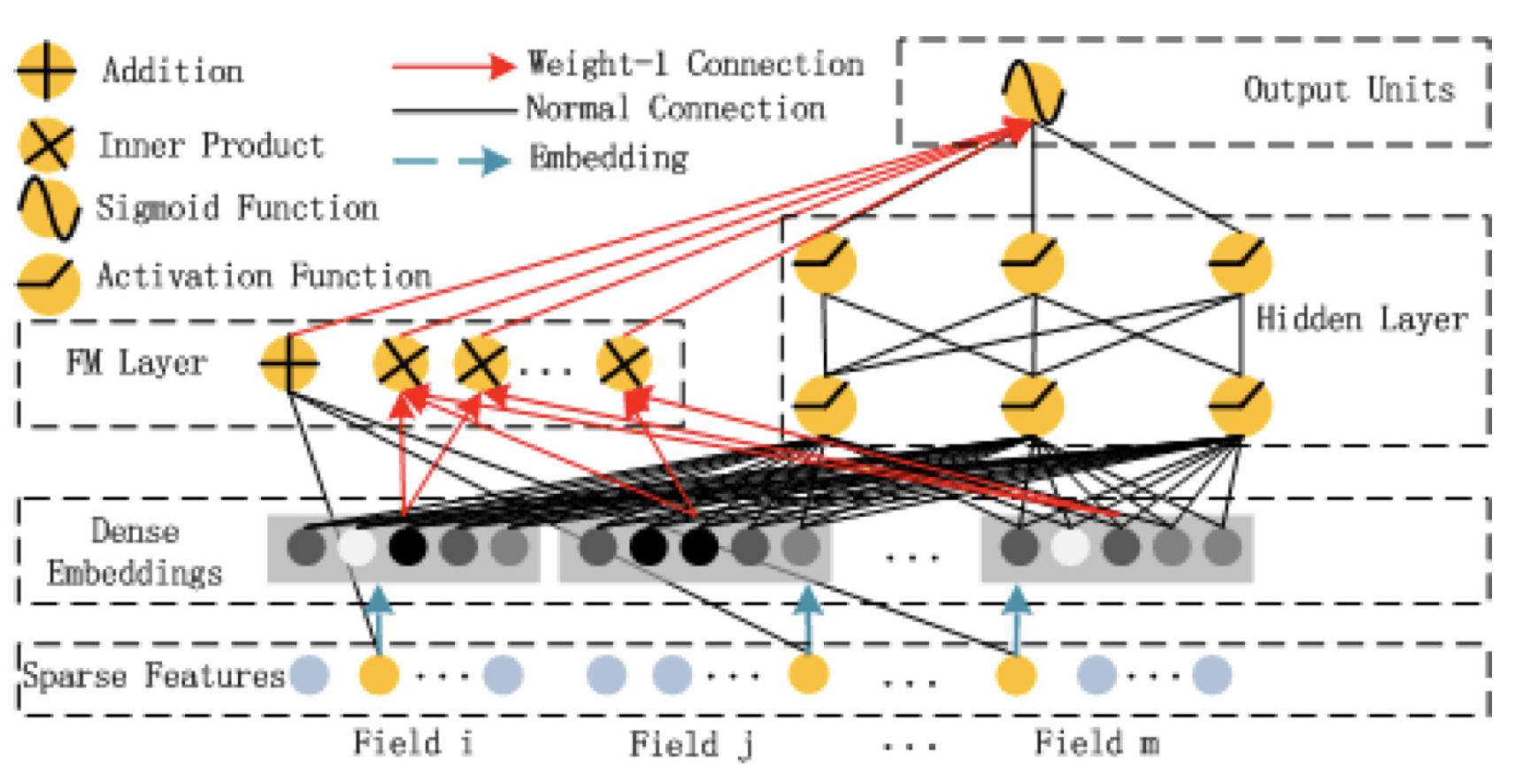}}
    \caption{Architecture of DeepFM model.}
    \label{fig:DeepFM}
\end{figure}

\noindent The architecture of DeepFM model is as figure \ref{fig:DeepFM}.

\noindent FM part of DeepFM don't need the feature engineering for inputs but it only realizes first-order and second-order feature interactions because of the parameter number. Suppose we have $m$ fields and embedding dimension is $k$, second-order feature interaction needs $O(k \times m^2)$ parameters while third-order needs $O(k \times m^3)$. Deep \& Cross Net \cite{wang2017Deep_Cross} is proposed to solve the parameters' rapid growth problem.


\subsubsection{Model Ensemble}
\noindent The concept of model ensemble is like the concept of wisdom of the crowd. We can take the advantage of a bunch independent models. To be more clear, adopt each one's good points and avoid its shortcomings. A list of common ensemble methods\cite{EnsembleMethods}:
\begin{enumerate}
    \item Vote
    \item Linear blending / weighted blending
    \item Bootstrap aggregation (Bagging)
    \item Stacking
\end{enumerate}

\subsubsection{Optimization Algorithms} \label{sec:optimization}
\noindent We can use one frame to explain all optimization algorithms since all those algorithms can be simplified into those four steps\cite{Optimization-frame}.

\begin{algorithm}[H]
    \begin{algorithmic}
    \STATE{Model parameters: $\theta \in \mathbb{R}^{d}$}
    \STATE{Target function: $J(\theta)$}
    \STATE{Learning rate: $\alpha$}
    \FOR{Each epoch $t$}
        \STATE{(1) Calculate target function's gradients over model current parameters:}
            \STATE{\qquad $g_{t}=\nabla J\left(\theta_{t}\right)$}
        \STATE{(2) Calculate 1st and 2nd order momentum based on past gradients:}
            \STATE{\qquad $m_{t}=\phi\left(g_{1}, g_{2}, \cdots, g_{t}\right) ; V_{t}=\psi\left(g_{1}, g_{2}, \cdots, g_{t}\right)$}
        \STATE{(3) Calculate the gradient should be eliminated:}
            \STATE{\qquad $\eta_{t}=\alpha \cdot m_{t} / (\sqrt{V_{t}} + \epsilon)$}
            \STATE{\qquad $\epsilon$ is a minimal value to protect denominator from zero, usually set $\epsilon=1e-8$}
        \STATE{(4) Update parameters:}
            \STATE{\qquad $\theta_{t+1}=\theta_{t}-\eta_{t}$}
    \ENDFOR
    \end{algorithmic}
    \caption{Outline of general optimization algorithms}
    \label{alg:optimization}
\end{algorithm}

\newpage

\bibliography{references.bib}

\end{document}